\documentclass[aps,pre,amsfonts,superscriptaddress,byrevtex]{revtex4}

\usepackage{graphicx}
\begin{document}
\begin{flushright}
{CAB-trans/04058}
\end{flushright}
\title{Spatio-temporal patterns driven by autocatalytic internal reaction
noise}
\author{David Hochberg}
\email{hochberg@laeff.esa.es} \homepage[]{http://www.cab.inta.es}
\affiliation{Centro de Astrobiolog\'{\i}a (CSIC-INTA), Ctra.
Ajalvir Km. 4, 28850 Torrej\'{o}n de Ardoz, Madrid, Spain}
\author{M.-P. Zorzano}
\email{zorzanomp@inta.es} \affiliation{Centro de
Astrobiolog\'{\i}a (CSIC-INTA), Ctra. Ajalvir Km. 4, 28850
Torrej\'{o}n de Ardoz, Madrid, Spain}
\author{Federico Mor\'{a}n}
\email{fmoran@bio.ucm.es} \affiliation{Centro de
Astrobiolog\'{\i}a (CSIC-INTA), Ctra. Ajalvir Km. 4, 28850
Torrej\'{o}n de Ardoz, Madrid, Spain} \affiliation{Departamento de
Bioqu\'{\i}mica y Biolog\'{\i}a
Molecular, Facultad de Ciencias Qu\'{\i}micas \\
Universidad Complutense de Madrid, 28040 Madrid, Spain}

\date{\today}
\begin{abstract}
The influence that intrinsic local density fluctuations can have
on solutions of mean-field reaction-diffusion models is
investigated numerically by means of the spatial patterns arising
from two species that react and diffuse in the presence of strong
internal reaction noise. The dynamics of the Gray-Scott (GS) model
with constant external source is first cast in terms of a
continuum field theory representing the corresponding master
equation. We then derive a Langevin description of the field
theory and use these stochastic differential equations in our
simulations. The nature of the multiplicative noise is specified
exactly without recourse to assumptions and turns out to be of the
same order as the reaction itself, and thus cannot be treated as a
small perturbation. Many of the complex patterns obtained in the
absence of noise for the GS model are completely obliterated by
these strong internal fluctuations, but we find novel spatial
patterns induced by this reaction noise in regions of parameter
space that otherwise correspond to homogeneous solutions when
fluctuations are not included.
\end{abstract}

\maketitle

\section{\label{sec:intro}Introduction}

General interest in the spatio-temporal pattern formation problem
stems from its wide application to self-organization phenomena in
fields as diverse as physics and chemistry \cite{CrossHohenberg},
to biology \cite{Camazine} and materials science \cite{Walgraef}.
One of the simplest models of biochemical relevance leading to
spatial and temporal patterns when diffusion is included is that
due to Gray and Scott (GS) \cite{Gray83}. Numerical simulations of
the deterministic GS system have revealed a rich set of strikingly
complex and irregular patterns \cite{Pearson}.  In these mean
field approximations of chemical species that diffuse and react,
the fluctuations are completely ignored. It is well known that if
the spacial dimensionality $d$ of the system is smaller than an
certain upper critical dimension $d_c$, the intrinsic fluctuations
play an important role in the late time asymptotic behavior and
the results obtained from the mean field equations are not correct
\cite{KangRedner}. Fluctuations can also influence the dynamics on
local spatial and temporal scales \cite{ZHM}. Indeed, it is well
established nowadays that noise can lead to an unsuspected variety
of dynamical effects. Far from being merely a perturbation to
idealized deterministic behavior or regarded as a bothersome
source of randomness or structural disruption, noise can induce
counterintuitive dynamical changes, two examples of which include
noise induced transitions \cite{Horsthemke} and stochastic
resonance \cite{Gammaitoni}.

In view of these considerations, and regarding the patterns
obtained in the mean-field approximation, it is important to
understand how fluctuations affect the stability of an established
spatial pattern and in what way do the deterministic and
stochastic effects compete. Fortunately, it is possible to include
systematically the effects of microscopic density fluctuations in
such systems by starting with the corresponding master equation,
representing this stochastic process by second-quantized bosonic
operators and then passing to a path integral representation to
map the system onto a continuum field theory \cite{Doi,Peliti}. In
many cases, primarily for two-body reactions, this field theory
can be mapped \emph{exactly} onto a Langevin equation description
in which the noise is completely and rigorously specified
\cite{LeeCardy}. In other cases, primarily for three-body
reactions to which the GS system belongs, the exact continuum
field theory can be approximated by nonlinear Langevin equations
provided the higher-order field theory vertices are truncated.
Since mean-field models of pattern formation are generally of the
reaction-diffusion type
\cite{CrossHohenberg,Camazine,Walgraef,Murray}, it is useful to
employ the Langevin description for handling the fluctuations, as
this allows for a direct comparison with the results obtained
within the naive mean field approximation.

In this paper we use the standard mapping of the master equation
to a stochastic field theory and use the latter to obtain a set of
approximate nonlinear Langevin equations in order to be able to
assess the nature and influence that internal reaction noise has
on the spatial patterns obtained from the purely deterministic GS
model.

The remainder of this paper is organized as follows. In Sec
{\ref{sec:model} we introduce the chemical reactions defining the
Gray-Scott model and derive a field theoretic description of these
reactions by means of the Doi-Peliti formalism \cite{Doi, Peliti}.
Once we obtain the continuum action, we derive an approximate
Langevin equation description of the GS model. The advantage of
this is that the noise properties are specified automatically and
indicate how the mean field reaction-diffusion equations must be
modified to take into account properly the (unavoidable) internal
density fluctuations. The ensuing noise is real and
multiplicative, and in magnitude is of the same order as the GS
reaction itself. In Sec \ref{sec:numsim} we present results of the
numerical simulations of the Langevin equations derived in Sec
{\ref{sec:model} and assess the impact that strong multiplicative
noise has on the subsequent evolution of spatially localized
structures in two dimensions. Conclusions are drawn in Sec
\ref{sec:concl}.

\section{\label{sec:model}The $U+2V \rightarrow 3V$ reaction}
The Gray-Scott model \cite{Gray83}, is a variant of the
autocatalytic Selkov model of glycolysis, corresponding to the
following chemical reactions:
\begin{eqnarray}\label{reaction}
U + 2 V &\stackrel{\lambda}{\to}& 3 V, \nonumber\\
V &\stackrel{\mu}{\to}& P, \nonumber\\
U &\stackrel{\nu}{\to}& Q, \nonumber\\
  &\stackrel{u_0}{\to}& U.
\end{eqnarray}
$V(x,t)$ and $U(x,t)$ represent the concentrations of the chemical
species $U$ and $V$, and are functions of $d$-dimensional space
$x$ and time $t$. $\lambda$ is the reaction rate, $P$ and $Q$ are
inert products, $\mu$ is the decay rate of $V$ and $\nu$ is the
decay rate of $U$ and $u_0$ is the constant feed rate.  A
non-equilibrium constraint is represented by a feed term for $U$.
The rate at which $U$ is supplied is positive if the concentration
of $U$ drops below an equilibrium value and negative if it exceeds
it. The equilibrium $U$ concentration is $u_0/\nu$, where $u_0$ is
the feed rate constant. The chemical species $U$ and $V$ can
diffuse with independent diffusion constants $D_u$ and $D_v$. All
the model parameters are positive.

\subsection{\label{sec:master}Continuous time master equation}

Our starting point is the continuous time master equation
describing the above reactions (\ref{reaction}) on a
$d$-dimensional hypercubic lattice, allowing multiple occupancy
per site. Consider the $U$ and $V$ particles moving diffusively on
a lattice of spacing $l$ and some probability of decaying, and of
reacting whenever they meet on a lattice site. Let
$P(\{m\},\{n\};t)$ be the probability to find the particle
configuration $\{m\},\{n\}$ at time $t$. The sets $\{m\} =
(m_1,m_2, \ldots, m_N)$ and $\{n\} = (n_1,n_2, \ldots, n_N)$
describe the occupation numbers of the $V$ and $U$ particles on
each lattice site $i$, respectively. The appropriate master
equation is given by
\begin{eqnarray}\label{ctmaster}
&&\frac{\partial}{\partial t}P(\{m\},\{n\};t) =
\frac{D_v}{l^2}\sum_{(i,j)} \{(m_j + 1)P(\ldots,m_i -1,m_j +
1,\ldots;t) -m_iP \} \nonumber \\
&+& \frac{D_u}{l^2}\sum_{(i,j)} \{(n_j + 1)P(\ldots,n_i -1,n_j +
1,\ldots;t) -n_iP \} \nonumber \\
&+& \frac{\lambda}{2} \sum_i\{(m_i -1)(m_i -2)(n_i +
1)P(\ldots,m_i - 1,\ldots,n_i + 1, \ldots; t) -m_i(m_i -
1)n_iP\}\nonumber \\
&+& \mu \sum_i \{(m_i + 1)P(\ldots,m_i + 1,\ldots;t)
-m_iP\}\nonumber \\
&+& \nu \sum_i \{(n_i + 1)P(\ldots,n_i + 1,\ldots;t)
-n_iP\}\nonumber \\
&+& u_0 \sum_i \{P(\ldots,m_i + 1,\ldots;t) - P\}.
\end{eqnarray}
This equation describes the evolution of $P$ in time. A given
configuration can change due to one of six independent processes:
by the diffusion of $V$ particles (first line of (\ref{ctmaster}))
where $D_v$ is the diffusion constant; by the diffusion of the $U$
particles (second line) where $D_u$ is the diffusion constant. It
will also change when two $V$ particles meet a $U$ particle to
produce another $V$ particle, with rate $\lambda$ (third line of
(\ref{ctmaster})); and when a $V$ particle or $U$ particle decay
spontaneously with rate $\mu$ and $\nu$, respectively (fourth and
fifth lines). Finally, the probability changes due to the constant
source of $U$ particles, with feed rate $u_0$ (sixth line). In the
diffusive terms the symbol $(i,j)$ indicates summing over sites
$i$ and their nearest neighbors $j$.

\subsection{\label{sec:hamiltonian}Mapping to bosonic field theory}

This master equation ({\ref{ctmaster}) can be mapped to a
second-quantized description following a procedure developed by
Doi \cite{Doi}. Briefly, we introduce annihilation and creation
operators $a$ and $a^\dag$ for $V$ and $b$ and $b^\dag$ for $U$ at
each lattice site, with the commutation relations
$[a_i,a_j^{\dag}] = \delta_{ij}$ and $[b_i,b_j^{\dag}] =
\delta_{ij}$. The vacuum state satisfies $a_i|0\rangle =
b_i|0\rangle = 0$. We then define the time-dependent state vector
\begin{equation}\label{wavefunction}
|\Psi(t)\rangle = \sum_{\{m\},\{n\}}P(\{m\},\{n\},t) \prod_i(
{a}_i^\dag)^{m_i}( {b}_i^\dag)^{n_i}|0\rangle.
\end{equation}
The master equation can be written as a Schr\"{o}dinger-like
equation
\begin{equation}\label{schrodinger}
-\frac{\partial |\Psi(t)\rangle}{\partial t} =
 {H}|\Psi(t)\rangle,
\end{equation}
where the lattice hamiltonian or time-evolution operator is a
function of $a,a^\dag,b,b^\dag$ and is given by
\begin{eqnarray}\label{hamiltonian}
 {H} &=& \frac{D_v}{l^2}\sum_{(i,j)}( {a_i}^\dag -
 {a_j}^\dag)( {a_i} -  {a_j} ) +
\frac{D_u}{l^2}\sum_{(i,j)}( {b_i}^\dag -
 {b_j}^\dag)( {b_i} -  {b_j} )\nonumber \\
&-&\frac{\lambda}{2}\sum_i[( {a_i}^\dag)^3  {a_i}^2 {b_i}
- ( {a_i}^\dag)^2  {a_i}^2  {b_i}^\dag  {b_i}]\nonumber \\
&+& \nu \sum_i( {b_i}^\dag -1) {b_i} + \mu \sum_i( {a_i}^\dag -1)
{a_i} + u_0\sum_i(1 -  {b_i}^\dag).
\end{eqnarray}
This has the formal solution $|\Psi(t)\rangle = \exp(-
{H}t)|\Psi(0)\rangle$.

Finally, this second quantized bosonic
operator(\ref{hamiltonian})is mapped onto a continuum field
theory. This procedure is now standard and we refer to
\cite{Peliti} for further details . In our case, for the GS
system, we end up with the following path integral
\begin{equation}\label{pathintegral}
U(\tau,0) = \int \mathcal{D}a \mathcal{D}\bar{a} \mathcal{D}b
\mathcal{D}\bar{b}\, e^{-S[a,\bar{a},b,\bar{b}]},
\end{equation}
over the continuous fields $a(x,t),\bar a(x,t), b(x,t), \bar
b(x,t)$ where the action $S$ is given by
\begin{equation}\label{action}
S = \int d^dx \int_0^{\tau} dt [\bar{a}\partial_t a + D_v\nabla
\bar{a} \nabla a + \bar{b}\partial_t b + D_u\nabla \bar{b} \nabla
b + \mu(\bar{a} - 1)a + \nu(\bar{b} - 1)b -u_0(\bar{b} - 1)
-\frac{\lambda}{2}(\bar{a}^3a^2b - \bar{a}^2a^2\bar{b}b) ].
\end{equation}
We have omitted terms related to the initial state. Aside from
taking the continuum limit, the derivation of this action is
exact, and in particular, no assumptions regarding the precise
form of the noise are required.

\subsection{\label{sec:Langevin}Approximate Langevin equation description}

For the final step we perform the shift $\bar{a} = 1 + a^*$ and
$\bar{b} = 1 + b^*$ on the action $S$ and obtain
\begin{eqnarray}\label{shiftaction}
S  = \int d^dx \int_0^{\tau} dt \Big( a^*\big(\partial_t a -
D_v\nabla^2 + \mu a  -\frac{\lambda}{2}a^2b \big) &+&
b^*\big(\partial_t b - D_u\nabla^2 b + \nu b - u_0 +
\frac{\lambda}{2}a^2b \big)  \nonumber \\
&-& \frac{\lambda}{2}a^2b[\underbrace{2{a^*}^2 -2a^*b^*} + {a^*}^3
- {a^*}^2b^*] \Big).
\end{eqnarray}

We will represent the quadratic terms in $a^*, b^*$ (indicated
with the underbrace) by an integration over Gaussian noise terms,
which will allow us to then integrate out the conjugate fields if
we ignore the terms cubic in these conjugate fields. Doing so, we
derive an approximate Langevin description of the exact field
theory in (\ref{action}). To carry this out explicitly, we note
that
\begin{equation}\label{identity}
e^{(\lambda a^2b({a^*}^2 - a^*b^*))} \approx \int \mathcal{D}\xi
\mathcal{D}\eta \, P(\xi,\eta)\,e^{(a^*\xi + b^* \eta)},
\end{equation}
where the noise functions $\xi,\eta$ are distributed according to
a double Gaussian as
\begin{equation}
P(\xi,\eta) = \exp\Big(-(\eta,\xi)A \left(\begin{array}{c}
  \eta \\
   \xi \\
\end{array} \right)
\Big),
\end{equation}
and where the (inverse) matrix $A$ of noise-noise correlation
functions is
\begin{equation}\label{matrix}
A = \frac{1}{2\lambda a^2 b}
\left(%
\begin{array}{cc}
  0 & 1 \\
  1 & 2 \\
\end{array}%
\right)\qquad A^{-1} =
\left(%
\begin{array}{cc}
  \langle \eta \eta \rangle & \langle \eta \xi \rangle \\
  \langle \xi \eta \rangle & \langle \xi \xi \rangle \\
\end{array}%
\right).
\end{equation}
Integrating out the conjugate fields $a^*$ and $b^*$ from the
functional integral (\ref{shiftaction}) then leads to the pair of
coupled nonlinear Langevin equations
\begin{eqnarray}\label{Langevin}
\partial_t a(x,t) &=& D_v\nabla^2 a(x,t) -\mu a(x,t) +
\frac{\lambda}{2}a(x,t)^2b(x,t) + \xi(x,t) \nonumber \\
\partial_t b(x,t) &=& D_u\nabla^2 b(x,t) -\nu b(x,t) -
\frac{\lambda}{2}a(x,t)^2b(x,t) + u_0 + \eta(x,t),
\end{eqnarray}
with positive noise correlations that can be read off directly
from (\ref{matrix})
\begin{eqnarray}\label{noisecorr}
\langle\xi(x,t)\rangle &=& \langle \eta(x,t) \rangle = 0 \nonumber
\\
\langle \xi(x,t)\xi(x',t')\rangle &=& \lambda a(x,t)^2b(x,t) \,
\delta^d(x-x')\delta(t-t') \nonumber \\
\langle \xi(x,t)\eta(x',t')\rangle &=& 2\lambda a(x,t)^2b(x,t)
\, \delta^d(x-x')\delta(t-t') \nonumber \\
\langle \eta(x,t)\eta(x',t')\rangle &=& 0.
\end{eqnarray}
Thus the multiplicative reaction noise is \emph{real}, a point
well worth mentioning since \textit{imaginary} noise terms are
known to arise in effective Langevin descriptions of diffusion
limited reactions \cite{Howard}. The mathematical structure of the
noise correlations in (\ref{noisecorr}) merits some comment. We
note that this equation establishes a relation between the moments
of the noise sources and the values of the fluctuating
concentration fields. Strictly speaking, the noise cumulants
should depend on the moments of the fields. Thus, for example, by
employing a so called ``curtailed characteristic functional", van
Kampen was able to exactly compute the cumulants of the noise
source for a spatially independent Markov process
\cite{vanKampen}. The apparent inconsistency in (\ref{noisecorr})
is common in the literature. We emphasize that this is not an
artifact of the approximation used here: similar relations (i.e.,
equating averaged to fluctuating variables) arise even when the
functional integration over the conjugate fields can be performed
exactly \cite{Howard}. There are two ways out of this apparent
inconsistency: on the one hand, we note that the noise cumulants
are proportional to delta functions which limit the effects of the
fluctuations to the single point $x = x'$ and single time $t =
t'$. Alternatively, in using the identity in (\ref{identity}), one
always has the freedom to define the matrix $A$ of noise
correlations (\ref{matrix}) as being strictly constant, at the
expense of shifting (the square-root of) the  field-dependent
prefactor into the resultant Langevin equation. If we follow this
option, the noise comes out strictly white and the noise
correlations are mathematically consistent. This in fact is the
option we employ below when we come to consider the numerical
simulations. Finally, we note that these Langevin equations
(\ref{Langevin}) reduce to the GS reaction-diffusion system in the
mean field approximation in which particles are uncorrelated.

\section{\label{sec:numsim}Numerical simulation}

Based on the microscopic master equation (\ref{ctmaster}) and the
field-theoretic action of the system (\ref{shiftaction}), we have
derived an approximate effective Langevin description
(\ref{Langevin}) for this chemical system where the statistical
properties of the internal noise terms (\ref{noisecorr}) have been
explicitly calculated. Notice that the unavoidable internal
reaction noise is multiplicative and its intensity is comparable
to that of the reaction terms. This problem can thus not be
analyzed by perturbation theory and must be treated numerically.
In the case of weak additive noise (Gaussian white noise and
colored Ornstein-Uhlenbeck noise) the stochastic system described
in Eq. (\ref{GSN}) has been investigated in detail
\cite{LHMP2,ZHM}. The patterns to which the system converged
changed drastically with small changes in the noise intensity.
Using the lowest order one-loop Renormalization Group (RG)
\cite{LHMP1}, we demonstrated that a weak additive noise induces a
modification in the parameters of the system. By combining
analytic and numerical work, we established an equivalence between
a sequence of patterns generated by varying the noise amplitude
but keeping all other parameters fixed, and a companion sequence
generated by keeping the noise fixed and varying (i.e.,
renormalizing) instead some of the model parameters according to
the predictions of the RG flow equations. In the deterministic
case, this reaction-diffusion system was numerically investigated
leading to a great variety of patterns in a rather small region of
the parameter space \cite{Pearson}. We investigate here this
region and its neighborhood when the reaction-diffusion system is
subjected to the intrinsically large amplitude internal reaction
noise.

The noise term in the $a$-equation has a field dependent
autocorrelation or strength. On the contrary, the noise in the
$b$-field equation has zero autocorrelation, i.e. it is a noise of
zero strength which we henceforth take as null in our numerical
studies. For real noise, we can identify the fields $a$ and $b$
with the particle densities $V$ and $U$, respectively.  Without
loss of generality, we redefine the noise $\xi \rightarrow
\sqrt{\lambda U }V \theta$ with $\theta$ a Gaussian white
(uncorrelated) zero-mean noise of unit strength. We thus consider
the following reaction-diffusion system subjected to
multiplicative noise in $d=2$ space dimensions:
\begin{eqnarray}\label{GSN}
\frac{\partial V}{\partial t}&=&\lambda U V ^{2} -\mu V + D_v
\nabla ^{2} V + \sqrt{\lambda U} V \theta (x,y,t)\\ \nonumber
\frac{\partial U}{\partial t}&=&u_0-\lambda  U V ^{2} -\nu U + D_u
\nabla ^{2}  U,
\end{eqnarray}
 with
\mbox{$<\theta _{v}(x,y,t)\theta
  _{v}(x',y',t')>$}=\ \mbox{$\delta(x-x')$}\ \mbox{$\delta(y-y')$}\ \mbox
{$\delta(t-t')$} and where   $\nabla ^{2}=\frac{\partial
^{2}}{\partial x^{2}}+\frac{\partial ^{2}}{\partial y^{2}}$. We
study this system for $\lambda=1$, $D_u=1$ and $D_v=0.5$ and
setting $u_0 = \nu$.

The numerical simulations of system evolution have been performed
using forward Euler integration of the finite-difference equations
following discretization of space and time in the stochastic
partial differential equations (\ref{GSN}). The spatial mesh
consists of a lattice of $256\times 256$ cells with cell size
$\Delta x=\Delta y= 2.18$ and periodic boundary conditions. The
noise has been discretized as well.  The system has been
numerically integrated up to $t=5000$ (with time step $\Delta
t=0.05$). After the transient time (roughly $t\approx 2000$,
depending on the exact system parameters and initial conditions),
during which the perturbation spread, the system went into an
asymptotic state.

For comparison with the deterministic case studied in
\cite{Pearson}, we have used the same $F,k$ coordinates which
correspond to $F=\nu$ and $k =\mu-\nu$. Following \cite{Pearson},
we first considered the time evolution of an initial perturbation
in the homogeneous trivial stable state of the reaction-diffusion
system. The initial conditions consisted of one localized square
pulse with ($U=0.5,V=0.25$) plus random Gaussian noise perturbing
the trivial steady state ($U=1,V=0$). The perturbing pulse
measured $22 \times 22$ cells, just wide enough to allow the
autocatalytic reaction to be locally self-sustaining.  In the
figures (a)-(e) and (R) in Fig. \ref{fig}, only the concentration
of the substrate $U$ is shown. When displayed in color, the blue
represents a concentration of $U$ between 0.2 and 0.4, where the
substrate is being depleted by the autocatalytic production of
$V$, yellow represents an intermediate concentration of roughly
0.8 and red represents the trivial steady state ($U=1,V=0$) where
all fluctuations cease entirely. None of the noise-free patterns
reported in \cite{Pearson} survive in this strong, stochastic
regime. However, besides the trivial time independent solution
shown in (R), ($U=1,V=0$), we have found non trivial spatial
patterns in a wider $k$ and $F$ region of parameter space than
that surveyed in \cite{Pearson}. In spite of the strong intrinsic
noise, the existence of the relatively ordered pattern (a) with
self-replicating moving globules is remarkable. These globules
consist of localized closed structures, in which the reactant
concentration differs from the surrounding concentration field. In
the interior of each of these units, in blue, there is a region
with sustained autocatalytic production of $V$ which is causing
the local depletion of the substrate $U$. This blue region
corresponds roughly to the state ($U=0.3$, $V=0.25$) depending on
the exact parameter values. The main difference between pattern
(a) and (b) is the ability of these structures to split into new
closed units, which is lost in pattern (b) leading to a merged
structure. For fixed $F=\nu$, as $k$ (or equivalently $\mu$, the
decay rate of the $V$ particles) is decreased, there is a smooth
transition from pattern (b)  ($k= 0.06$) through (e) ($k = 0.025$)
and then again to (R). Similar patterns can be found with
different $k$ and $F$. As $F$ is decreased the size of the
structures in the patterns increases, see for instance Fig.
\ref{fig1} and compare the patterns with the corresponding ones
shown in Fig. \ref{fig}.

In Fig. \ref{fig2} we present the parameter space mapping of the
patterns found. The solid-line separates two relevant regions of
the deterministic reaction-diffusion case: on the right-side of
the solid line there is a single trivial, spatially uniform state
(R) whereas on the left-side there are two spatially uniform
states (R and uniform blue). Both are linearly stable. In the
vicinity of this line, as $F$ is decreased, the uniform blue
states looses stability through a Hopf bifurcation leading to a
great variety of patterns. For a detailed analysis of the patterns
found in parameter space, for the time evolution of this initial
condition in the deterministic case, see \cite{Pearson}. Notice
that pattern (a) appears for a set of parameters that under the
deterministic case would have led to a trivial solution of the
type (R). On the contrary, the patterns found in the deterministic
reaction-diffusion case do not survive when the internal reaction
noise is taken into account. In particular, the uniform blue state
disappears. Therefore of the two uniform stable solutions of the
deterministic case, only the trivial one survives. In the trivial
red state $(U=1,V=0)$, the stochastic fluctuations, whose
amplitude is given by $\sqrt{\lambda U}V$, cease entirely and
hence this state is {\em inactive}. Whereas in the blue uniform
state ($U=0.3,V=0.25$) the non-vanishing fluctuations drive the
system away to one of the patterns shown. Furthermore, the range
in parameter space over which patterns can be found is greater in
the noisy case than in the deterministic one: the range is roughly
twice as wide in the $k$-range and approximately three times as
wide in the $F$-range.

For completeness we considered next the case of uniform,
unperturbed, initial conditions. As mentioned before, for the
parameter region on the left-side of the solid line in Fig.
\ref{fig2}, the blue state ($U=0.3,V=0.25$) is the non-trivial
stable solution of the \emph{noise-free} reaction-diffusion
system. In Fig. \ref{blue}, upper row, we show the time evolution
of this state for $F=0.05$ and $k=0.055$, i.e. within the region
where it should be stable in the noise-free case. However, due to
the reaction noise, the blue pattern evolves towards pattern (c),
like the uniform red state did under the influence of local
perturbation, see Fig. \ref{fig2}}. Therefore, the local density
fluctuations are strong enough to spontaneously form a pattern
also starting from the blue uniform stable state. If we set
$F=0.05$ and $k=0.0725$, which corresponds to the right-side of
the solid line, we find the evolution of this pattern converges to
the globular pattern (a), see the lower row of figures in Fig.
\ref{blue}. This is also remarkable since in the noise-free
reaction-diffusion case, spots cannot form spontaneously from a
uniform state.

Therefore, if we take into account the unavoidable intrinsic
reaction noise, the dynamics of the system can be completely
different, and in some cases, even richer than that of the
idealized noise-free case. We also found that if the noise
intensity of this multiplicative stochastic term is artificially
damped, we do recover all the complex patterns obtained for the
purely deterministic mean-field situation.

\section{\label{sec:concl}Discussion}

The standard (noise-free) GS reaction-diffusion equations presume
that there is particle diffusion due to the uncorrelated Brownian
motion of the molecules involved and that the reaction rate is
simply given by the product of the probability of finding two
molecules of autocatalyst and a molecule of substrate at the same
point. This approximation clearly neglects the existing
correlation between the molecules and the presence of microscopic
particle density fluctuations which cause these mean field rate
equations to break down.

Starting from the microscopic master equation, we have derived a
field-theoretic action of the GS reaction system and from there we
have deduced effective Langevin-type equations where the form of
the noise is specified precisely without any assumption. An
alternative but equivalent approach yielding a path integral
solution to the chemical master equation, and which dispenses with
creation and annihilation operators, is given by R. Kubo, K.
Matsuo and K. Kitahara \cite{KuboMatsuoKitahara}. The Langevin
equations derived here are approximate in that the multiplicative
noise appearing in (\ref{Langevin}) is Gaussian distributed.
Recall that the noise represents the terms in the action $S$
(\ref{shiftaction}) quadratic in the conjugate fields $a^*,b^*$.
The presence of cubic terms in these fields indicates that the
fluctuations are actually skewed (not symmetric), but there is
unfortunately no exact identity (i.e., in the spirit of
Hubbard-Stratanovich) allowing one to replace the cubic terms by
equivalent noise terms, as we did in (\ref{identity}).
Nevertheless, we note that the quadratic and cubic terms are all
proportional to the reaction term $\sim \lambda a^2 b$. We thus
expect that the noise in the putative exact Langevin description
has a strength comparable with that of the Gaussian approximation.
The internal reaction noise depends both on the density of
substrate and product, i.e. when either of them is zero there is
no reaction and therefore no noise either. The numerical solutions
in Fig 1 (a-d) and in Fig 3 correspond to \emph{active states},
i.e., since in these states, the internal fluctuations persist and
the asymptotic particle densities are finite. In fact,
fluctuations are always present in the GS model, since the
substrate $U$ is being constantly replenished at the rate $u_0
> 0$, and provided that $V$ is non-vanishing, the noise amplitude
$\sqrt{\lambda U}V$ is always positive and finite. Only for the
trivial state $U=1,V=0$, where the density of $V$ is vanishing, do
the stochastic fluctuations cease entirely and hence the state (R)
is \emph{inactive}.

The internal reaction noise is unavoidable and as strong as the
reaction term. We have demonstrated numerically that its influence
can dramatically change the dynamics of the system producing new
stable states in the reaction. In particular, we report on the
existence of globular replicating structures in the Langevin GS
reaction-diffusion system, with internal reaction noise, in a
region of parameter space which in the deterministic case was
expected to decay to the trivial, uniform solution.

Through this study-case of a chemical reaction system we have
provided a specific example where the evolution of the densities
depends strongly on microscopic fluctuations, and cannot be
derived from mean-field rate equations. This approach may also be
relevant in other chemical processes capable of generating
spatially organized structures, and in particular, in the case of
low spatial dimensionality.

\begin{acknowledgments}
We thank Marcel Vlad for a critical reading of the manuscript, for
useful discussions, and for bringing our attention to the papers
cited in \cite{KuboMatsuoKitahara,vanKampen}.  M.-P.Z.
acknowledges a fellowship provided by INTA for training in
astrobiology. The research of D.H. is supported in part by funds
from CSIC, Comunidad Aut\'onoma de Madrid and F.M. acknowledges
partial support from Grant \# BMC2003-06957 from MEC (Spain).
\end{acknowledgments}

\clearpage
\newpage

\begin{figure}[h]
\begin{center}
\begin{tabular}{cccccc}
  R &   a & b & c & d & e\\
\includegraphics[width=0.13 \textwidth]{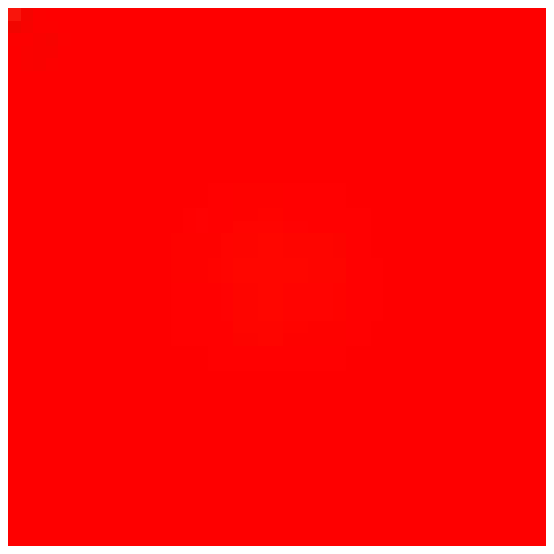} &
\includegraphics[width=0.13 \textwidth]{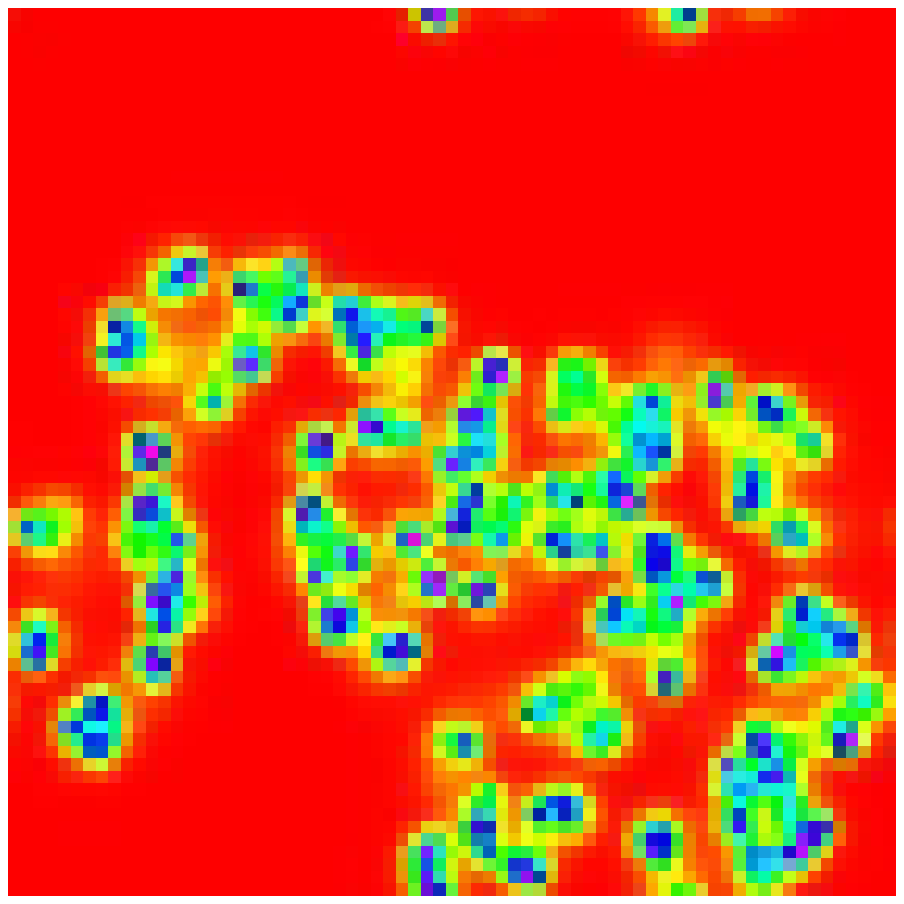} &
\includegraphics[width=0.13 \textwidth]{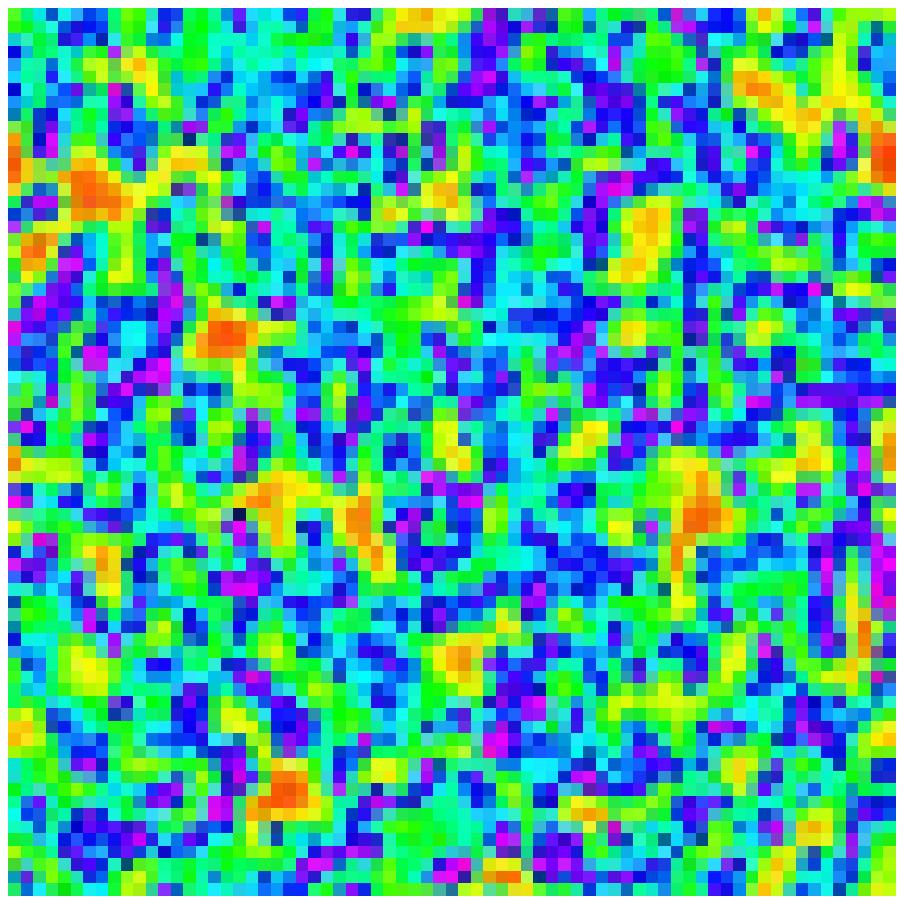} &
\includegraphics[width=0.13 \textwidth]{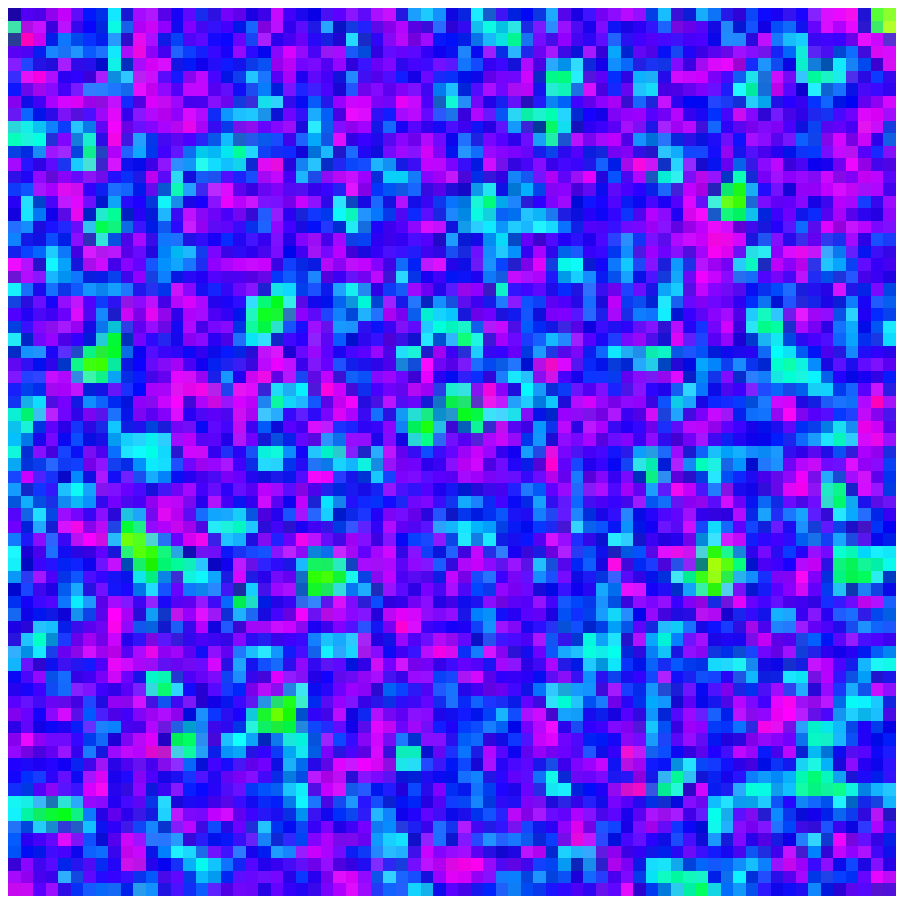} &
\includegraphics[width=0.13 \textwidth]{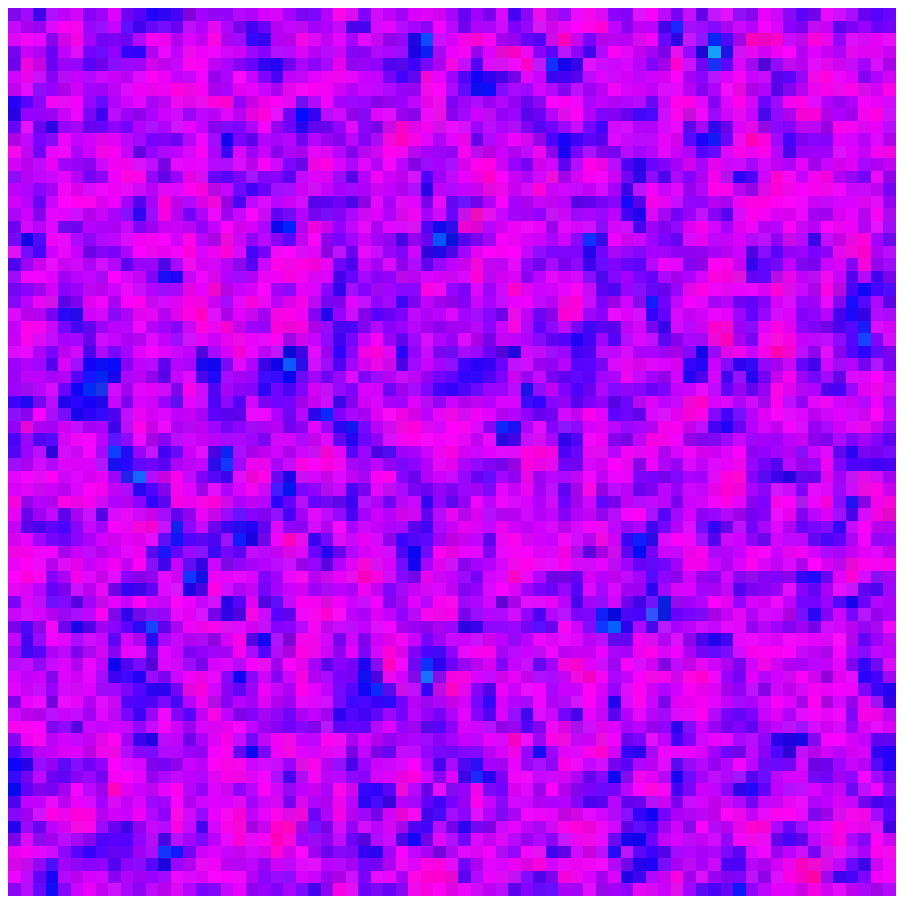} &
\includegraphics[width=0.13 \textwidth]{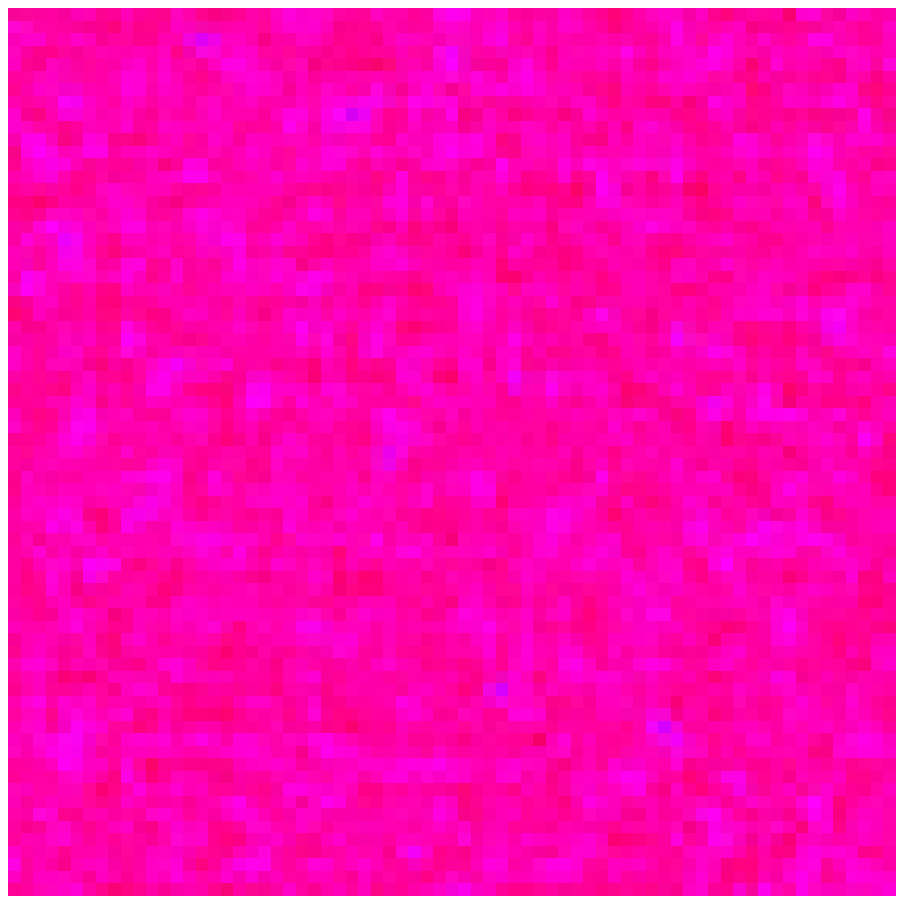}  \\
$k=0.08$ & $k=0.07$ & $k=0.06$ & $k=0.05$ & $k=0.04 $&  $k=0.03$
\end{tabular}
\caption{\label{fig}Reference patterns of the GS
reaction-diffusion system subject to internal reaction noise. The
initial condition was the uniform red, trivial state ($U=1,V=0$)
with a small localized, perturbing pulse (see text for details).
Concentration of field $U(x,y,t)$ is displayed at $t=5000$ for the
parameter range $F=\nu=0.025$, $k=\mu-\nu=[0.025:0.08]$.}
\end{center}
\end{figure}
\clearpage
\newpage

\begin{figure}[h]
\begin{center}
\begin{tabular}{ccc}
 a & b & c \\
\includegraphics[width=0.13 \textwidth]{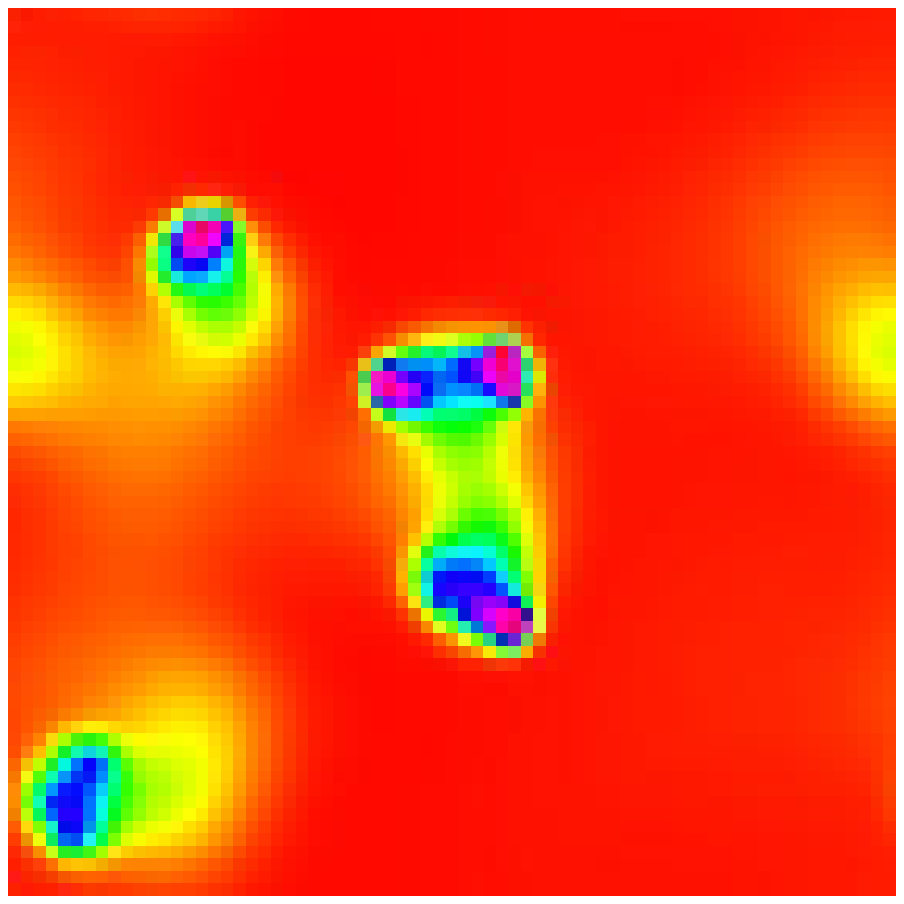} &
\includegraphics[width=0.13 \textwidth]{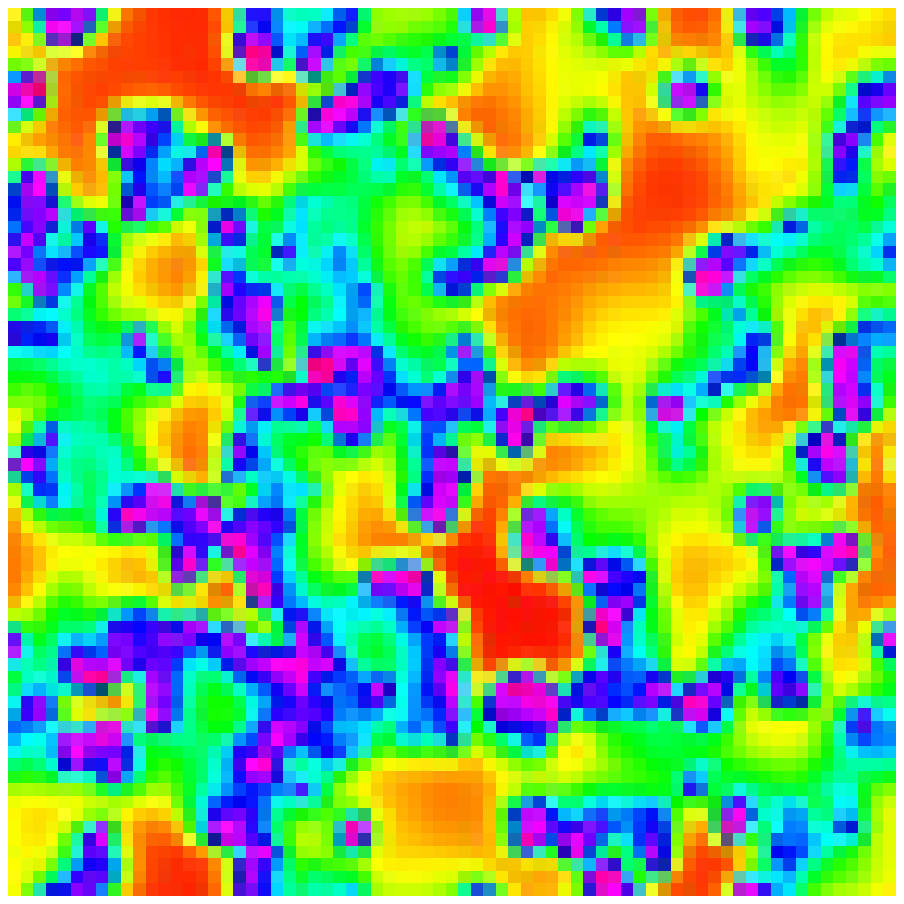}  &
\includegraphics[width=0.13 \textwidth]{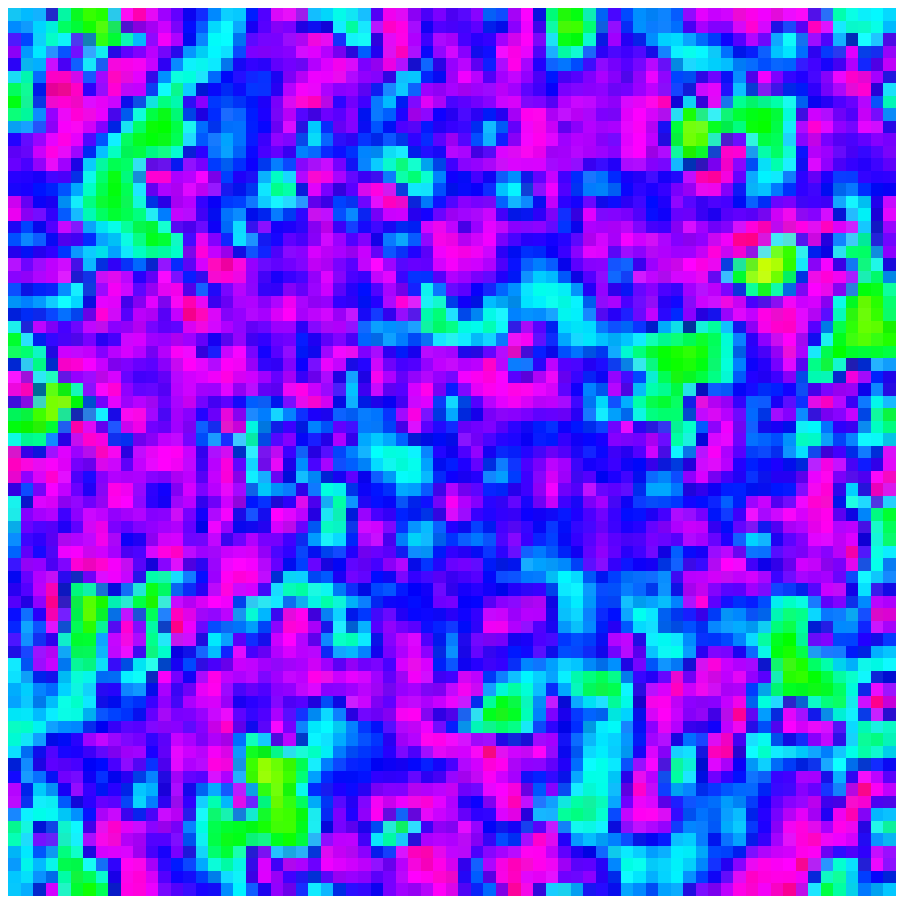}   \\
  $F=0.0025$ & $F=0.01$ & $F=0.01$
\end{tabular}
\caption{\label{fig1} The concentration of the field $U(x,y,t)$
displayed at $t=5000$ for (a)  $F$=0.0025, $k=0.05$ (b) $F$=0.01,
$k=0.05$ (c) $F$=0.01, $k=0.0375$. As $F$ is decreased the size of
the structures in the patterns increases. The patterns are
designated with the corresponding letter of the reference case in
Fig. \ref{fig}, which have a greater value of $F$. Notice that the
size of the square region depicted in the figures is the same.}
\end{center}
\end{figure}
\clearpage
\newpage

\begin{figure}
\begin{center}
\includegraphics[width=.8\textwidth]{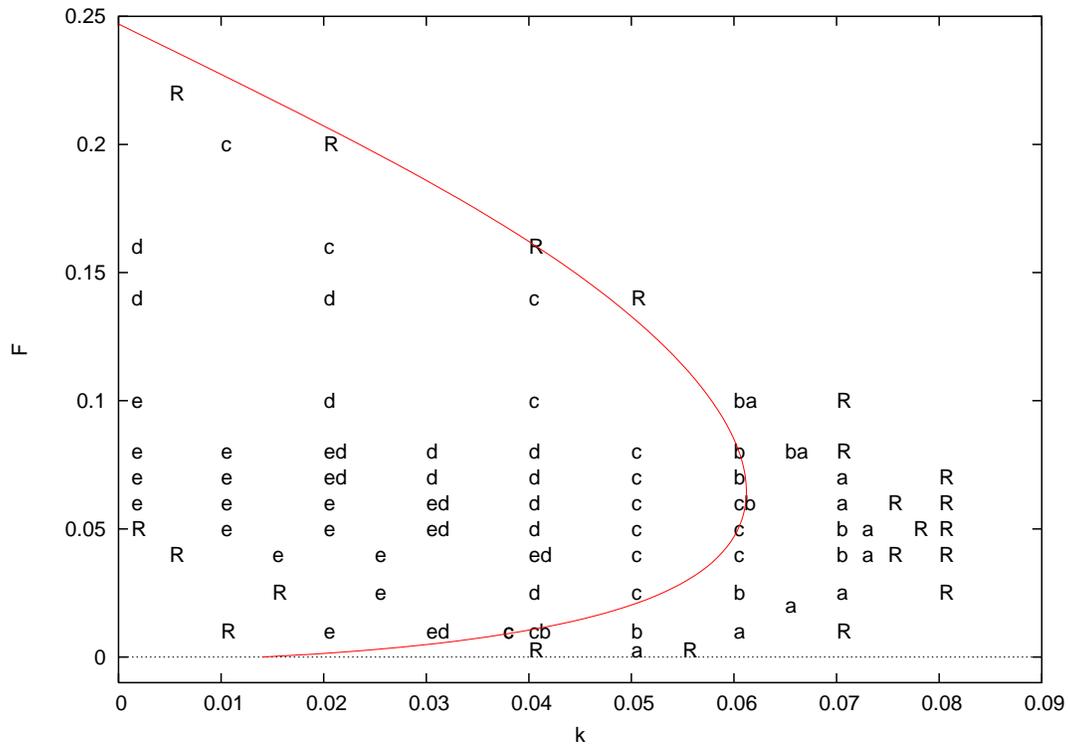}
\caption{\label{fig2}Parameter space diagram with $k =\mu-\nu$ and
$F =\nu$. The letters indicate the location where similar patterns
to the reference ones in Fig. (\ref{fig}) were found. A
transitional pattern between two reference cases is designated by
a pair of two corresponding letters (e.g., ed and  cb). R
indicates that the system evolved to the \emph{inactive} uniform
trivial, red, state. See the text for an explanation of the
solid-line.}
\end{center}
\end{figure}
\clearpage
\newpage

\begin{figure}[h]
\begin{center}
\begin{tabular}{cccc}
    B & i & ii & c \\
\includegraphics[width=0.13 \textwidth]{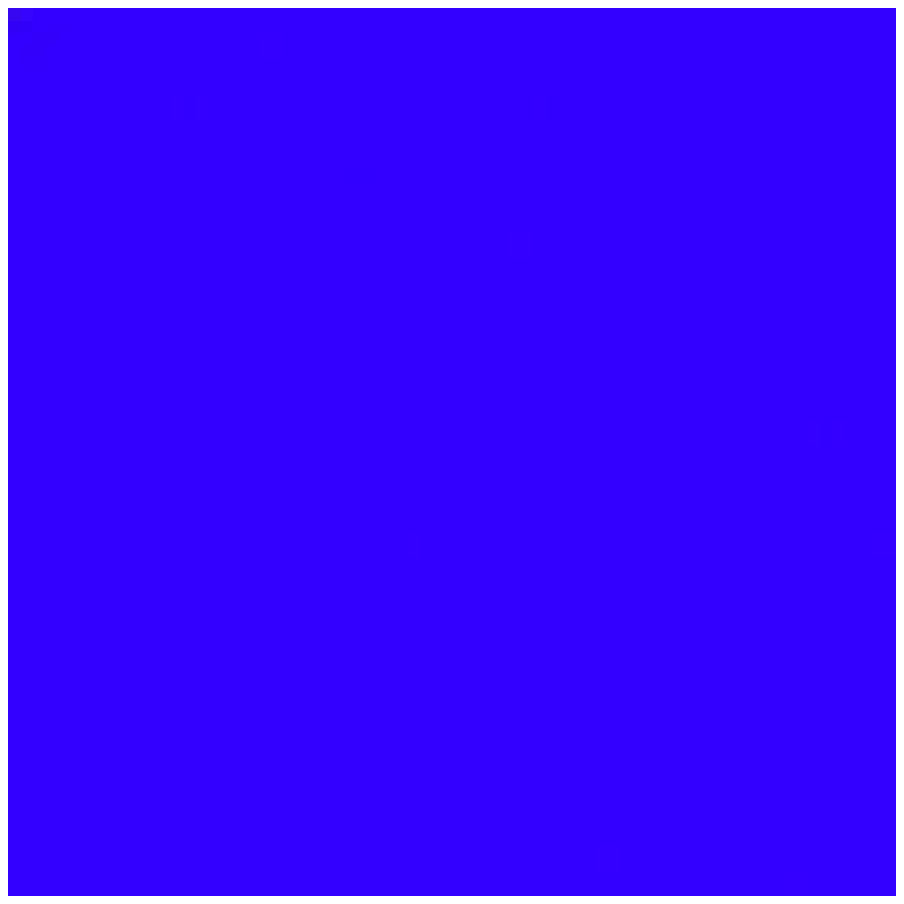} &
\includegraphics[width=0.13 \textwidth]{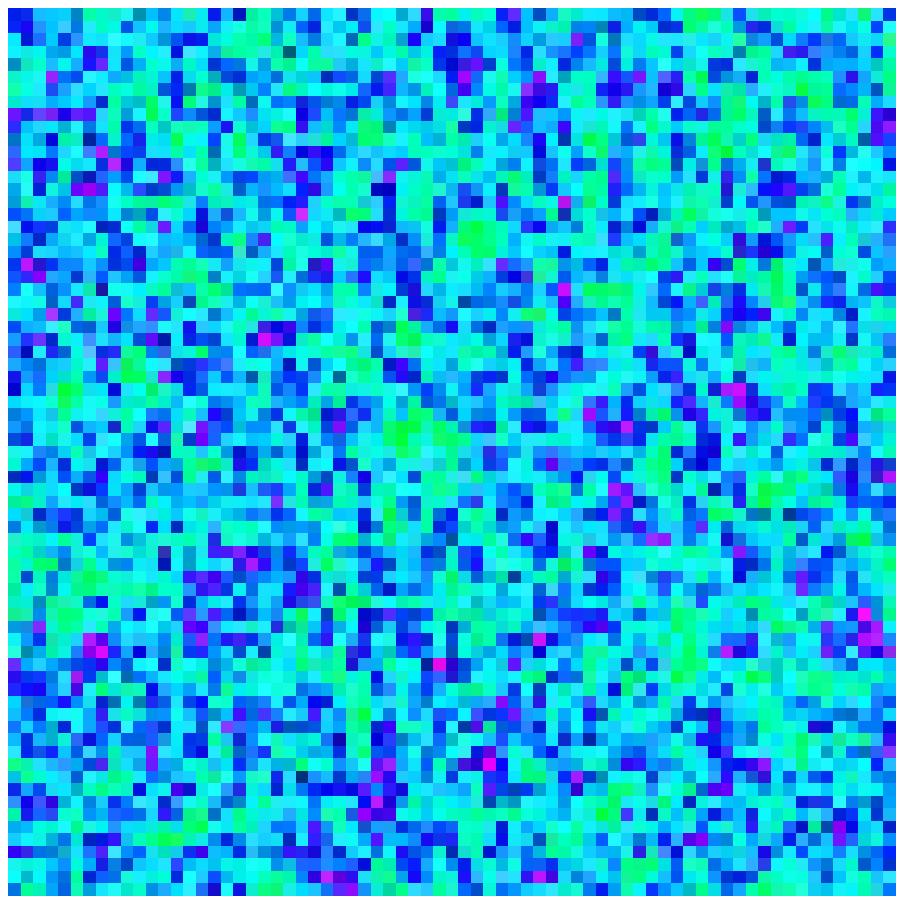} &
\includegraphics[width=0.13 \textwidth]{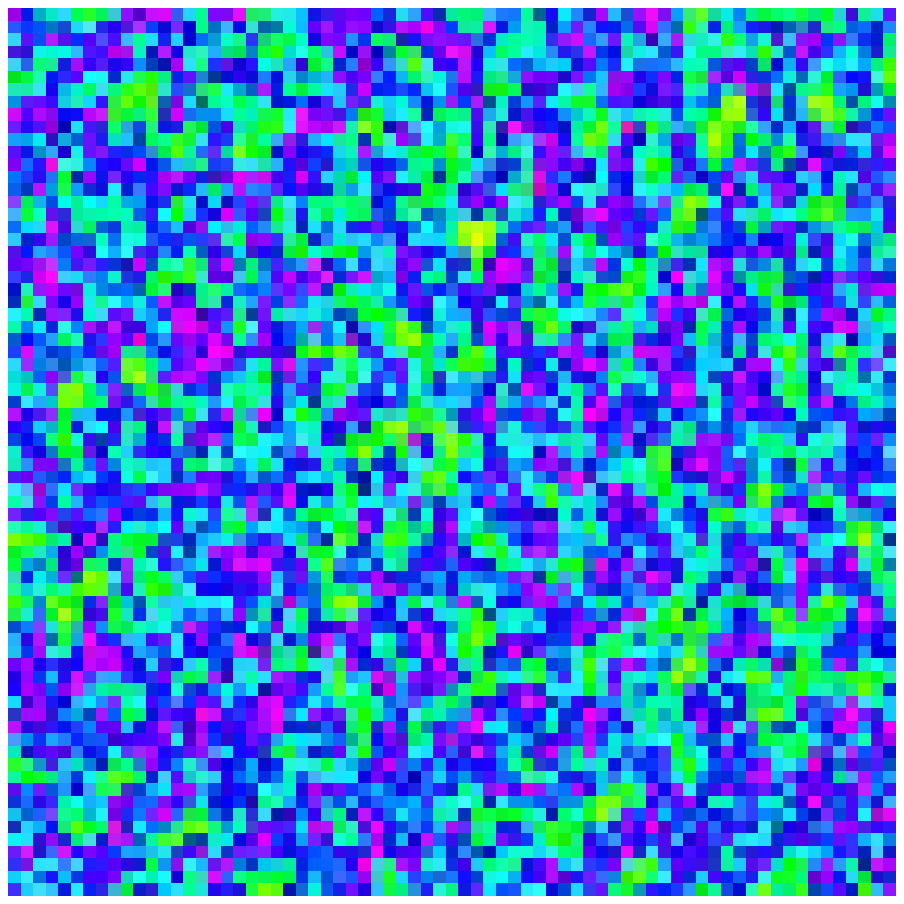} &
\includegraphics[width=0.13 \textwidth]{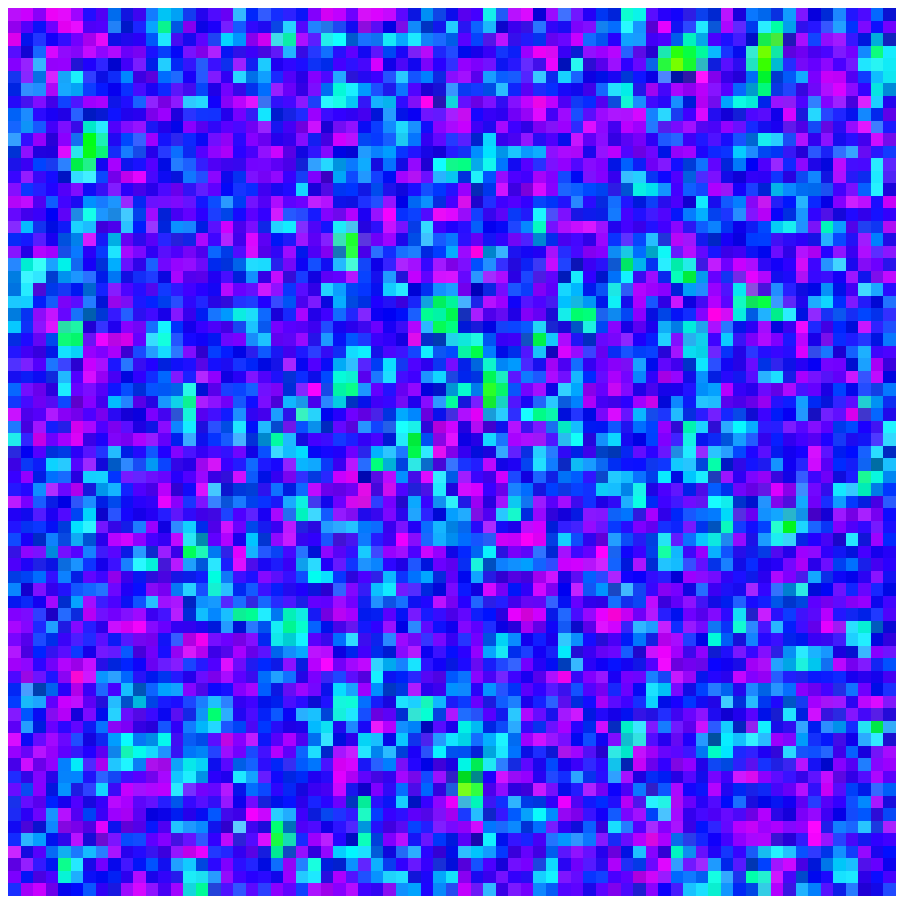}  \\
 $(F,k)=(0.05,0.055)$\ \ $t=0$  &  $t=20$ & $t=40$ & $t=600$
\end{tabular}
\begin{tabular}{cccc}
    B & I & II & a \\
\includegraphics[width=0.13 \textwidth]{IC_blue.eps} &
\includegraphics[width=0.13 \textwidth]{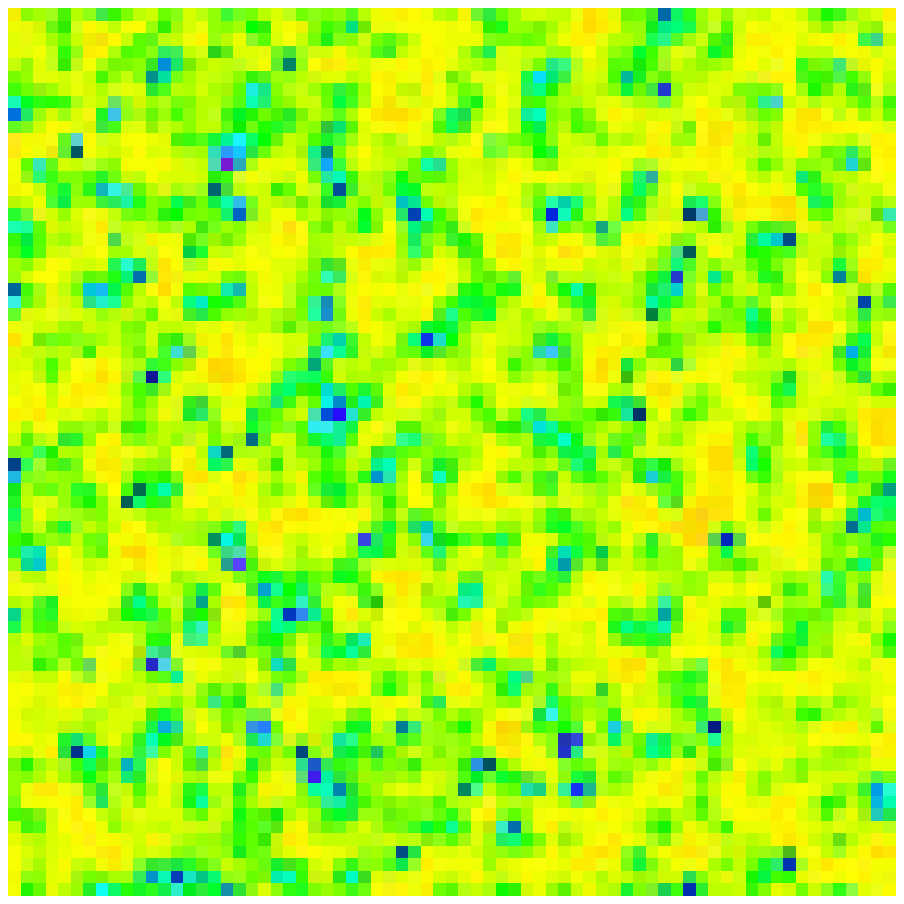} &
\includegraphics[width=0.13 \textwidth]{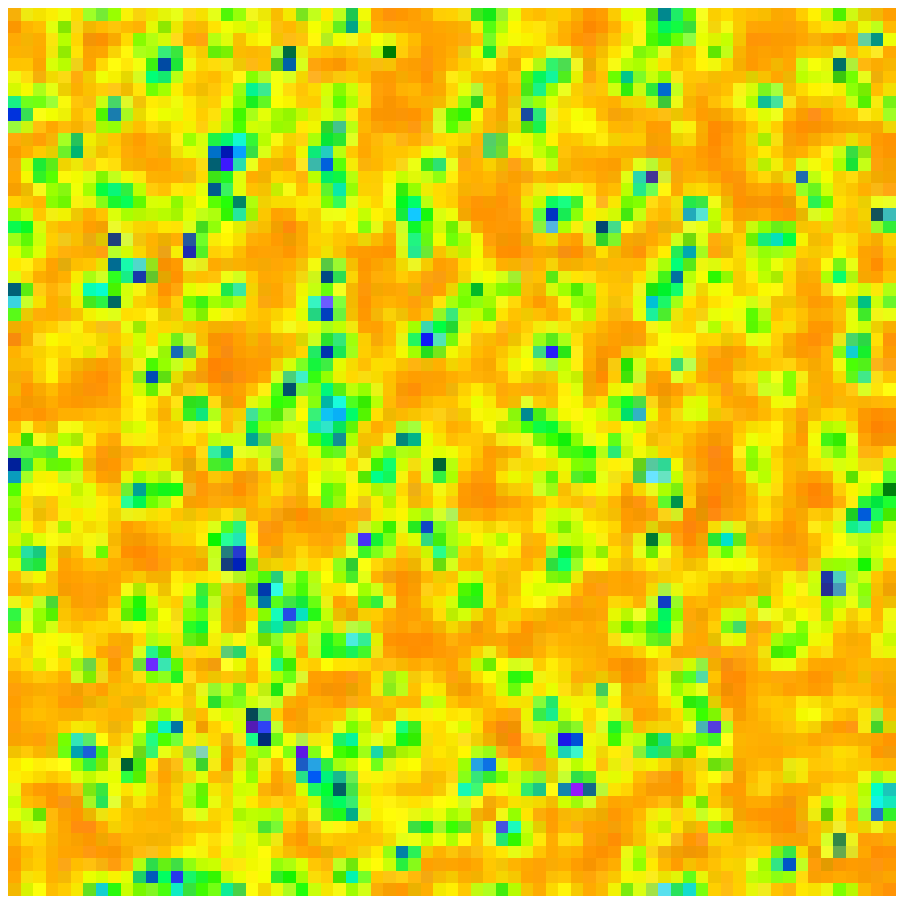} &
\includegraphics[width=0.13 \textwidth]{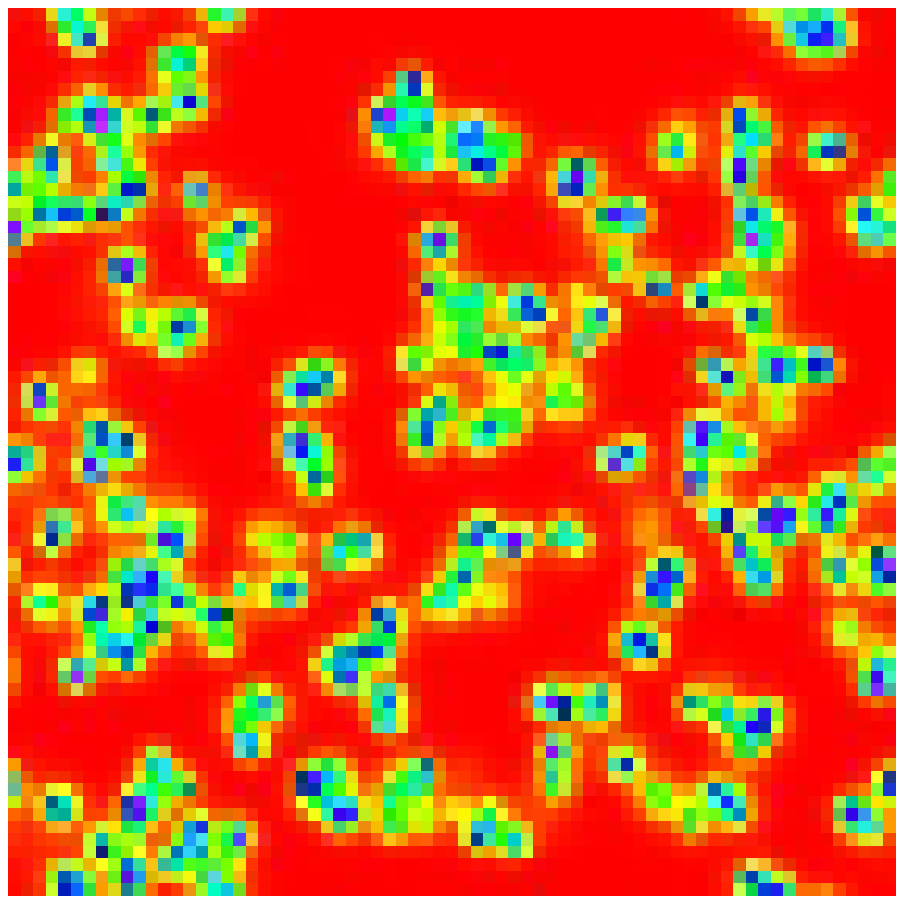}  \\
$(F,k)=(0.05,0.0725)$\ \ $t=0$ &  $t=40$ & $t=50$ & $t=3500$
\end{tabular}
\caption{\label{blue}Time evolution of the homogeneous initial
condition ($U=0.3,V=0.25$): (B)-(i)-(ii)-(c) for $k=0.055$ and
$F=0.05$, and (B)-(I)-(II)-(a) for $k=0.075$ and $F=0.05$.}
\end{center}
\end{figure}

\end{document}